\documentclass[11pt,a4paper]{article}

\usepackage{amsmath,amssymb,amsfonts}
\usepackage{bm}
\usepackage{physics}
\usepackage{geometry}
\usepackage{hyperref}
\usepackage{setspace}
\usepackage{graphicx}
\usepackage{xcolor}
\usepackage{cite}   
\usepackage{doi}
\newcommand{\be}{\begin{equation}}
\newcommand{\ee}{\end{equation}}
\begin{document}

\title{\textbf{Krylov Complexity for Time-Dependent Hamiltonians}}

\author{
\textbf{Amin Faraji Astaneh}$^{a,b}$\footnote{faraji@sharif.ir}
\quad and \quad
\textbf{Parsa Kafashi}$^{a}$\footnote{parsa.kafashi@physics.sharif.edu}
}

\date{}

\maketitle

\begin{center}
\emph{$^{a}$ Department of Physics, Sharif University of Technology,\\
P.O. Box 11155-9161, Tehran, Iran}
\end{center}

\begin{center}
\emph{$^{b}$ Research Center for High Energy Physics,\\
Department of Physics, Sharif University of Technology,\\
P.O. Box 11155-9161, Tehran, Iran}
\end{center}

\thispagestyle{empty}

\begin{abstract}
We investigate Krylov spread complexity for states evolving under time-dependent Hamiltonians. For periodically driven systems, we formulate the problem within Floquet theory and show how the Magnus expansion provides a systematic approximation when the Floquet Hamiltonian is not available in closed form. We then extend this framework beyond periodic driving and demonstrate that, in addition to the globally truncated Magnus expansion, a piecewise Magnus expansion provides a reliable method when the global expansion loses convergence or accuracy. Our results provide practical tools for analyzing complexity growth in a broad class of time-dependent quantum systems.
\end{abstract}

\newpage
\tableofcontents
\newpage

\section{Introduction}

Krylov complexity, together with its state-based counterpart often referred to as spread complexity, has recently emerged as an important bottom-up measure of quantum complexity; for reviews, see \cite{Nandy2025Quantum,Rabinovici2025Krylov}. It quantifies how an initially simple state or operator explores Hilbert space under time evolution. While Krylov complexity has been extensively studied for time-independent Hamiltonians \cite{Barbon:2019wsy,Dymarsky:2021bjq,Takahashi:2025kct,Caputa:2023rhh,Caputa:2021vub,Astaneh:2025spread,Alishahiha:2022nhe,Imani:2025etp,Baume:2026jyt}, its extension to time-dependent Hamiltonians introduces additional conceptual and technical challenges.

For a time-dependent Hamiltonian \(H(t)\), the time-evolution operator is given by
\begin{equation}
U(t,t_0)
=
\mathcal{T}
\exp\left[
-i\int_{t_0}^{t} H(s)\,ds
\right],
\end{equation}
where \(\mathcal{T}\) denotes time ordering.\footnote{Throughout this work, we set \(\hbar=1\).}
The time-ordering operation is necessary because, in general,
\begin{equation}
[H(t_1),H(t_2)]\neq 0.
\end{equation}
This noncommutativity makes it difficult to apply the standard Krylov construction directly, since that construction is naturally formulated for time-independent generators. The main goal of this paper is to propose practical methods for overcoming this subtlety and computing the Krylov complexity of states for such time--dependent Hamiltonians.

Periodically driven systems provide an important setting in which time-dependent dynamics can sometimes be recast in terms of an effective time-independent generator through Floquet theory; for a review, see \cite{Sato2025Floquet, Bukov2015Floquet}. This makes it possible to define a stroboscopic notion of Krylov complexity using the Floquet Hamiltonian. However, in many cases the Floquet Hamiltonian cannot be obtained exactly in closed form. In such situations, perturbative methods such as the Magnus expansion, and in the periodic setting the Floquet--Magnus expansion, provide systematic approximations to the effective Hamiltonian. These methods are especially useful in high-frequency regimes, where the driving period is short compared with the intrinsic timescales of the system.

Beyond periodic driving, one may also consider general time-dependent Hamiltonians for which no global Floquet description exists. For such systems, a global Magnus expansion may be useful at short times, but it can lose convergence or accuracy over long time intervals. To address this limitation, we develop a piecewise Magnus approach. The idea is to divide the full time interval into short subintervals, approximate the evolution on each subinterval by a local effective Hamiltonian, and reconstruct the full time evolution as an ordered product of these local evolutions. This allows the Krylov complexity to be extracted from a sequence of controlled local approximations.

The structure of this paper is as follows. In Section~2, we review periodically driven systems and the basic ingredients of Floquet theory. In Section~3, we review the Magnus expansion and its specialization to the Floquet--Magnus expansion. In Section~4, we apply this framework to periodically driven two-level systems, considering both circularly and linearly polarized drives, the rotating-wave approximation, the high-frequency Floquet--Magnus expansion, and the associated truncation-error bounds. In Section~5, we extend the discussion to non-periodic time dependence by introducing the piecewise Magnus expansion for a harmonic oscillator with time-dependent frequency; we first treat the sudden quench as a benchmark and then apply the method to smooth finite-time, or soft, quenches. Section~6 contains our conclusions and discusses the relation of our approach to other Krylov-based treatments of time-dependent Hamiltonians. The appendices provide a review of Krylov complexity and present details of some calculations and proofs.

\section{Periodically Driven Systems and Floquet Theory}

A suitable and manageable starting point for our survey is periodically driven systems, a prominent class of time-dependent quantum systems governed by Hamiltonians satisfying
\begin{equation}
H(t+T)=H(t)\, ,
\end{equation}
where \(T\) is the period of the drive. For such systems, Floquet theory provides a powerful framework for describing the dynamics.

The central object is the one-period evolution operator,
$
U(t_0+T,t_0)
$.
Since this operator is unitary, it can be written as the exponential of a Hermitian operator:
\begin{equation}
U(t_0+T,t_0)=e^{-iH_FT}\, .
\end{equation}
This equation defines the Floquet Hamiltonian \(H_F\)\footnote{It is important to note that the Floquet Hamiltonian is not unique. This is because it is defined through the logarithm of a unitary operator, and the logarithm is multi-valued. If \(\epsilon_\alpha\) is a quasienergy, then
\begin{equation*}
\epsilon_\alpha
\sim
\epsilon_\alpha+\frac{2\pi m_\alpha}{T},
\qquad
m_\alpha\in\mathbb{Z}.
\end{equation*}
Therefore, different choices of the logarithm correspond to different quasienergy branches. These choices lead to Floquet Hamiltonians that generate the same one-period evolution operator. In what follows, we choose a convenient branch whenever a specific expression for \(H_F\) is required. Terms proportional to the identity shift all quasienergies uniformly and contribute only an overall phase; hence, they do not affect transition probabilities or the Krylov complexity considered here.}.

At stroboscopic times
\begin{equation}
t_n=t_0+nT\,,
\qquad
n\in\mathbb{Z}\,,
\end{equation}
the evolution is
\begin{equation}
U(t_0+nT,t_0)
=
\left[U(t_0+T,t_0)\right]^n
=
e^{-inH_F T}.
\end{equation}
Thus, at stroboscopic times, the periodically driven system evolves exactly as though it were governed by the static Hamiltonian \(H_F\).

The full time evolution also contains intra-period dynamics, or micromotion. In general, Floquet theory allows one to write
\begin{equation}
U(t,t_0)=P(t)\,e^{-iH_F(t-t_0)}\,P^\dagger(t_0)\, ,
\end{equation}
where \(P(t)\) is a periodic micromotion operator satisfying
\begin{equation}
P(t+T)=P(t).
\end{equation}
Equivalently, one may write
\begin{equation}
P(t)=e^{-iK_F(t)}\, ,
\end{equation}
where \(K_F(t)\) is the kick operator. In the Floquet gauge, one chooses \(P(t_0)=\mathbb I\), so the evolution operator becomes
\begin{equation}
U(t,t_0)=P(t)\,e^{-iH_F(t-t_0)}\, .
\end{equation}
Since \(P(t)\) is periodic, this choice implies
\begin{equation}
P(t_0+nT)=\mathbb{I}\, .
\end{equation}
Thus, at stroboscopic times, the micromotion disappears and the dynamics is governed solely by \(H_F\).

The distinction between stroboscopic and non-stroboscopic evolution is important for the interpretation of Krylov complexity. At stroboscopic times, and in the Floquet gauge satisfying \(P(t_0)=\mathbb{I}\), the dynamics is fully generated by \(H_F\). Away from stroboscopic times, however, the micromotion operator contributes nontrivially:
\begin{equation}
\ket{\psi(t)}
=
P(t)e^{-iH_F(t-t_0)}\ket{\psi(t_0)}.
\end{equation}
Therefore, even when the stroboscopic Floquet Hamiltonian does not generate transitions in a chosen basis, the full time-dependent state may still exhibit nonzero instantaneous complexity due to intra-period micromotion. 
\section{The Magnus and Floquet--Magnus Expansions}

It is not always possible to compute the Floquet Hamiltonian exactly. Moreover, not every time-dependent Hamiltonian is periodic, and therefore Floquet theory is not always directly applicable. In such situations, the Magnus expansion provides a systematic method for approximating the time-evolution operator \cite{Magnus1954}; see also the review \cite{Blanes2009Magnus}.
The Magnus expansion handles the time-ordering problem by writing
\begin{equation}
U(t,0)
=
\mathcal{T}\exp\left[-i\int_0^t H(s)\,ds\right]
=
e^{\Omega(t)}\, ,
\end{equation}
where
\begin{equation}
\Omega(t)=\Omega_1(t)+\Omega_2(t)+\Omega_3(t)+\cdots\, ,
\end{equation}
and the first few terms in the expansion are
\begin{align}
\Omega_1(t)
&=
-i\int_0^ H(t_1)\,dt_1,
\\
\Omega_(t)
&=
-\frac{1}{2}
\int_0^t dt_1
\int_0^{t_1} dt_2\,
[H(t_1),H(t_2)],
\\
\Omega_3(t)
&=
\frac{i}{6}
\int_0^t dt_1
\int_0^{t_1} dt_2
\int_0^{t_2} dt_3
\Big(
[H(t_1),[H(t_2),H(t_3)]]
+
[H(t_3),[H(t_2),H(t_1)]]
\Big).
\end{align}
At leading order, the time-dependent Hamiltonian is therefore replaced by its
average over the time interval. This average-Hamiltonian approximation is not always accurate. Higher-order
Magnus terms contain nested commutators of the Hamiltonian at different times,
and these terms become important when
$
[H(t_1),H(t_2)]\neq 0 
$. Thus, the reliability of the leading approximation depends on the size of these
commutators and on the length of the time interval considered.
Truncating the expansion at order \(m\), one obtains
\begin{equation}
U(t,0)\approx e^{\Omega^{(m)}(t)}\, ,
\qquad
\Omega^{(m)}(t)=\sum_{k=1}^{m}\Omega_k(t)\, .
\end{equation}
A corresponding effective Hamiltonian can be defined by
\begin{equation}
H_{\mathrm{eff}}^{(m)}
=
\frac{i}{t}\Omega^{(m)}(t)\, ,
\end{equation}
so that
\begin{equation}
U(t,0)\approx e^{-iH_{\mathrm{eff}}^{(m)}t}\, .
\end{equation}

The accuracy of the Magnus expansion is not guaranteed for arbitrarily long times. Since the expansion contains time integrals of nested commutators, higher-order terms may become increasingly important as the total evolution time grows. Thus, a finite-order truncation may be reliable at short times but gradually lose accuracy at longer times due to the accumulation of errors. In periodically driven systems, this limitation can manifest itself through secular terms, resonances, or heating effects that are not captured by a low-order approximation.

A commonly used sufficient condition for convergence is \cite{Moan2008Convergence}
\begin{equation}
\int_0^t \norm{H(s)}\,ds < \pi .
\end{equation}
Stronger bounds have also been obtained in \cite{Apel2025Magnus}.
Although this condition is not necessary in all cases, it illustrates that the Magnus expansion is naturally suited to short-time evolution, high-frequency driving, and regimes where the relevant commutators are small. This observation motivates the piecewise Magnus expansion, which we introduce later together with a more detailed discussion of its truncation error.

For periodically driven systems, one often applies the Magnus expansion over a single period:
\begin{equation}
U(T,0)=e^{-iH_FT}.
\end{equation}
In this case, the Floquet Hamiltonian is approximated by
\begin{equation}
H_F^{(m)}
=
\frac{i}{T}\Omega^{(m)}(T).
\end{equation}
This is the Floquet--Magnus expansion.
Given this Floquet Hamiltonian, Krylov complexity can be constructed using an effective time-independent generator, thereby bypassing the time-ordering difficulties that arise in the direct treatment of the original time-dependent Hamiltonian. 
\section{Two-Level Periodically Driven Systems}

We now apply the preceding framework to driven two-level systems. In many cases,
the Floquet Hamiltonian can be written, up to an irrelevant identity
term,\footnote{A more general two-level Floquet Hamiltonian may contain an
additional term \(h_0\mathbb I\). This term contributes only an overall phase to
the time evolution and therefore does not affect transition probabilities or
the spread complexity considered here. We therefore omit it.}
as
\begin{equation}\label{two level Floquet Hamiltonian}
H_F
=
h_x\sigma_x
+
h_y\sigma_y
+
h_z\sigma_z\, .
\end{equation}
We choose the initial state to be the \(\sigma_z\) eigenstate
\begin{equation}
\ket{\psi(0)}=\ket{0}\, ,
\qquad
\sigma_z\ket{0}=\ket{0}\, .
\end{equation}
The orthogonal state is denoted by \(\ket{1}\), with
\begin{equation}
\sigma_z\ket{1}=-\ket{1}\, .
\end{equation}
For a two-dimensional Krylov basis generated from \(\ket{0}\), we may take
\begin{equation}
\{\ket{K_0},\ket{K_1}\}
=
\{\ket{0},\ket{1}\}\, .
\end{equation}
If
\begin{equation}
\ket{\psi(t)}
=
\phi_0(t)\ket{0}
+
\phi_1(t)\ket{1}\, ,
\end{equation}
then
\begin{equation}
C_K(t)
=
|\phi_1(t)|^2\, .
\end{equation}
For the Hamiltonian in Eq.~\eqref{two level Floquet Hamiltonian} one finds
\begin{align}
\phi_0(t)
&=
\cos(ht)
-
i\frac{h_z}{h}\sin(ht),
\\
\phi_1(t)\, 
&=
-i\frac{h_x+ih_y}{h}\sin(ht)\,,
\end{align}
where
\begin{equation}
h\equiv\sqrt{h_x^2+h_y^2+h_z^2}\,.
\end{equation}
Therefore,
\begin{equation}\label{generalCK}
C_K(t)
=
\frac{h_x^2+h_y^2}{h^2}
\sin^2(ht)\, .
\end{equation}
At stroboscopic times, the stroboscopic complexity is obtained by substituting \(t=nT\).

\subsection{Circularly Polarized Drive}

Consider the circularly polarized two-level Hamiltonian
\begin{equation}
H_{\mathrm{circ}}(t)
=
\frac{\omega_0}{2}\sigma_z
+
\frac{\Omega_0}{2}
\left(
\sigma_x\cos\omega t
+
\sigma_y\sin\omega t
\right).
\end{equation}
To remove the explicit time dependence, we transform to the rotating frame
\begin{equation}
R(t)=e^{-i\frac{\omega t}{2}\sigma_z}.
\end{equation}
Since \(R(0)=\mathbb I\) and \(R(T)=-\mathbb I\), with
\(T=2\pi/\omega\), this transformation changes the one-period evolution
operator only by an overall phase. This phase does not affect observables or
the spread complexity.

In the rotating frame,
\begin{equation}
\ket{\widetilde \psi(t)}
=
R^\dagger(t)\ket{\psi(t)},
\end{equation}
and the Hamiltonian becomes
\begin{equation}
\widetilde{H}_{\mathrm{circ}}
=
R^\dagger(t)H_{\mathrm{circ}}(t)R(t)
-
iR^\dagger(t)\dot{R}(t).
\end{equation}
Using the standard Pauli-matrix rotation identities, one obtains
\begin{equation}
\widetilde{H}_{\mathrm{circ}}
=
\frac{\Delta}{2}\sigma_z
+
\frac{\Omega_0}{2}\sigma_x,
\end{equation}
where
\begin{equation}
\Delta=\omega_0-\omega
\end{equation}
is the detuning.\footnote{Because \(R(T)=-\mathbb I\), the corresponding
lab-frame Floquet Hamiltonian may contain an additional identity term, such as
\((\omega/2)\mathbb I\), depending on the chosen quasienergy branch. This term
only contributes an overall phase and therefore does not affect transition
probabilities or the spread complexity considered here.}

Thus,
\begin{equation}
h_x=\frac{\Omega_0}{2},
\qquad
h_y=0,
\qquad
h_z=\frac{\Delta}{2}.
\end{equation}
Substitution into the general two-level result gives the stroboscopic
complexity
\begin{equation}
C_K^{\mathrm{circ}}(nT)
=
\frac{\Omega_0^2}{\Omega_0^2+\Delta^2}
\sin^2\left[
\frac{nT}{2}
\sqrt{\Omega_0^2+\Delta^2}
\right].
\end{equation}
\subsection{Linearly Polarized Drive}

We now consider the linearly polarized two-level Hamiltonian
\begin{equation}\label{eq:linpol-hamiltonian}
H_{\mathrm{lin}}(t)
=
\frac{\omega_0}{2}\sigma_z
+
\Omega_0\sigma_x\cos\omega t.
\end{equation}
This drive contains both co-rotating and counter-rotating components.
Therefore, unlike the circularly polarized case, a single rotating-frame
transformation cannot eliminate the full time dependence exactly.

\subsubsection{Near-Resonant Regime and the Rotating-Wave Approximation}

When the drive frequency is close to the level splitting,
$
\omega\approx \omega_0
$,
the perturbation can resonantly induce transitions between the two levels.
Then, one can decompose the drive into co-rotating and counter-rotating
contributions. In the rotating frame, the co-rotating term varies slowly near
resonance, whereas the counter-rotating term oscillates rapidly. In the
rotating-wave approximation (RWA), the latter is neglected.

This yields the effective Hamiltonian
\begin{equation}
\widetilde{H}_{\mathrm{RWA}}
=
\frac{\Delta}{2}\sigma_z
+
\frac{\Omega_0}{2}\sigma_x\, ,
\end{equation}
where $\Delta$ denotes the detuning introduced earlier. This is the same reduced effective Hamiltonian obtained for
the circularly polarized drive. Therefore, in the near-resonant RWA regime, the complexity is identical to that
obtained for the circularly polarized drive. The corresponding Rabi frequency is
\begin{equation}
\Omega_R
=
\sqrt{\Delta^2+\Omega_0^2}\, .
\end{equation}
On exact resonance, \(\Delta=0\), this reduces to
$
\Omega_R=\Omega_0
$.

\subsubsection{High-Frequency Regime and the Floquet--Magnus Expansion}\label{section4.2.2}

We now consider the high-frequency regime. As noted above, for the linearly
polarized Hamiltonian, unlike in the circularly polarized case, a single
conventional rotating-frame transformation cannot render the full Hamiltonian
time independent. One component can be made static, but the counter-rotating
component remains oscillatory. It is therefore natural to employ a
Floquet--Magnus expansion to construct an effective stroboscopic Hamiltonian.

A convenient starting point is to remove the transverse drive exactly by means
of the periodic dressing transformation
\begin{equation}
R(t)
=
\exp\!\left[
-i\frac{\Omega_0}{\omega}\sin(\omega t)\sigma_x
\right]\, .
\end{equation}
Since
\begin{equation}
R(0)=R(T)=\mathbb{I}\, ,
\end{equation}
this transformation does not change the stroboscopic evolution operator. In
the dressed frame the Hamiltonian becomes
\be
\widetilde{H}_{\mathrm{lin}}(t)=
\frac{\omega_0}{2}
\left[
\cos\!\left(
\frac{2\Omega_0}{\omega}\sin\omega t
\right)\sigma_z
+
\sin\!\left(
\frac{2\Omega_0}{\omega}\sin\omega t
\right)\sigma_y
\right]\, .
\label{eq:dressed-hamiltonian}
\ee
Thus, the explicit drive term is removed, but the noncommuting static term is
converted into a time-dependent operator.
The leading Floquet--Magnus contribution in the dressed frame is the time
average,
\be
H_{\mathrm{eff}}^{(1)}=
\frac{1}{T}\int_0^T \widetilde{H}_{\mathrm{lin}}(t)\,dt=
\frac{\omega_0}{2}
J_0\!\left(\frac{2\Omega_0}{\omega}\right)\sigma_z \, .
\ee
Expanding the Bessel function,
\begin{equation}
J_0\!\left(\frac{2\Omega_0}{\omega}\right)
=
1
-
\frac{\Omega_0^2}{\omega^2}
+
\frac{\Omega_0^4}{4\omega^4}
+\cdots .
\end{equation}
Thus, the Bessel factor resums an infinite subset of corrections in
\(\Omega_0/\omega\) already at leading order in the dressed-frame
average-Hamiltonian approximation.

We now include the next Magnus term. The second Magnus contribution to the
effective Hamiltonian is
\begin{equation}
H_{\mathrm{eff}}^{(2)}
=
-\frac{i}{2T}
\int_0^T dt_1
\int_0^{t_1} dt_2\,
[
\widetilde H_{\mathrm{lin}}(t_1),
\widetilde H_{\mathrm{lin}}(t_2)
].
\label{eq:second-magnus-general}
\end{equation}
A detailed derivation of this term is given in Appendix~\eqref{appB}. The final result is
\begin{equation}
H_{\mathrm{eff}}^{(2)}
=
-\frac{\pi\omega_0^2}{4\omega}
J_0\!\left(\frac{2\Omega_0}{\omega}\right)
\mathbf H_0\!\left(\frac{2\Omega_0}{\omega}\right)
\sigma_x .
\label{eq:heff-second-magnus-struve}
\end{equation}
For small \(\Omega_0/\omega\),
\begin{equation}
\mathbf H_0\!\left(\frac{2\Omega_0}{\omega}\right)
=
\frac{4\Omega_0}{\pi\omega}
+
O\!\left(\frac{\Omega_0^3}{\omega^3}\right),
\end{equation}
and therefore
\begin{equation}
H_{\mathrm{eff}}^{(2)}
=
-\frac{\omega_0^2\Omega_0}{\omega^2}\sigma_x
+
O\!\left(\frac{\omega_0^2\Omega_0^3}{\omega^4}\right).
\end{equation}
Thus, up to second order in the dressed-frame Magnus expansion,
\begin{equation}
h_z
=
\frac{\omega_0}{2}
J_0\!\left(\frac{2\Omega_0}{\omega}\right),
\label{eq:hz-definition}
\end{equation}
and
\begin{equation}
h_x
=
-\frac{\pi\omega_0^2}{4\omega}
J_0\!\left(\frac{2\Omega_0}{\omega}\right)
\mathbf H_0\!\left(\frac{2\Omega_0}{\omega}\right).
\label{eq:hx-definition}
\end{equation}
The sign of \(h_x\) depends on the sign convention chosen for the
\(\sigma_y\) term in the dressed-frame Hamiltonian, but the Krylov complexity
depends only on \(h_x^2\).

The Bessel-renormalized term in \(h_z\) is the leading average-Hamiltonian
result, while \(h_x\) is the first genuine commutator correction. This
transverse term is absent at the average-Hamiltonian level and is responsible
for weak transitions even when the initial state is an eigenstate of
\(\sigma_z\).

Let us emphasize again that, at stroboscopic times \(t=nT\), the dressing unitary satisfies
\(R(nT)=\mathbb{I}\). Hence the dressed-frame and lab-frame states coincide
stroboscopically within this Floquet gauge. Now one can use \eqref{generalCK} to obtain the complexity.
Away from the zeros of \(J_0\), and to leading order in \(\omega_0/\omega\),
this simplifies to\footnote{At zeros of the Bessel function, the leading average-Hamiltonian
quasienergy splitting vanishes. Interestingly, these zeros are the familiar
coherent-destruction-of-tunneling (CDT) points: the drive renormalizes the
effective tunneling matrix element by \(J_0(2\Omega_0/\omega)\), so that
\(J_0(2\Omega_0/\omega)=0\) dynamically quenches the leading tunneling
process \cite{Ashhab2007}. In the present second-order dressed-frame
truncation, both \(h_z\) and \(h_x\) are proportional to
\(J_0(2\Omega_0/\omega)\), so the stroboscopic dynamics is frozen through
this order. Residual dynamics near these points is determined by higher
Magnus corrections and micromotion effects.}
\begin{equation}
C_K(nT)
\approx
\frac{\pi^2\omega_0^2}{4\omega^2}
\mathbf H_0^2\!\left(\frac{2\Omega_0}{\omega}\right)
\sin^2\!\left[
\frac{\omega_0 nT}{2}
J_0\!\left(\frac{2\Omega_0}{\omega}\right)
\right].
\end{equation}
For \(\Omega_0/\omega\ll 1\), this further reduces to
\begin{equation}
C_K(nT)
\approx
\frac{4\omega_0^2\Omega_0^2}{\omega^4}
\sin^2\!\left[
\frac{\omega_0 nT}{2}
J_0\!\left(\frac{2\Omega_0}{\omega}\right)
\right].
\end{equation}

Finally, we emphasize that the transverse term
\eqref{eq:heff-second-magnus-struve} is Floquet-gauge dependent: in a different
choice of stroboscopic origin, part of this
correction may be shifted between the effective Hamiltonian and the micromotion
operator. The stroboscopic observables are, of course, gauge independent when
all orders are retained.

\begin{figure}[t]
    \centering
    \includegraphics[width=0.75\linewidth]{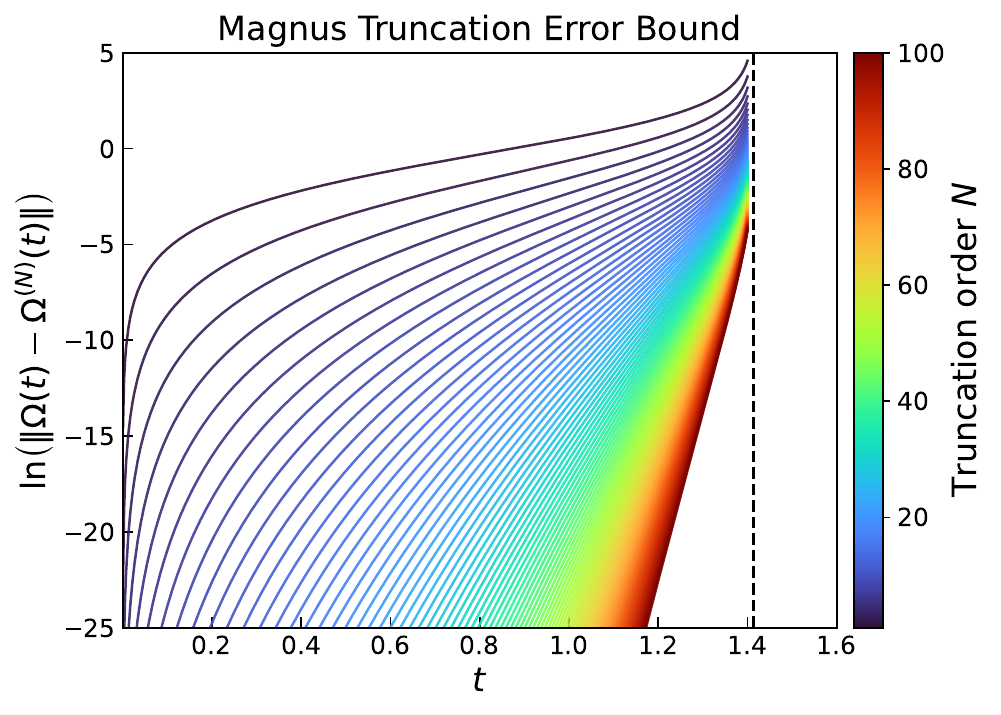}
    \caption{
    Magnus truncation error bound for the time-dependent Hamiltonian
    \eqref{eq:linpol-hamiltonian}. The figure shows the bound in \eqref{magnustruncationerror} as a function of time for
    truncation orders \(N=1\) to \(100\). The vertical dashed line indicates
    the convergence radius in time, beyond which the bound diverges. Higher
    truncation orders yield exponentially improved accuracy within the
    convergence regime.
    }
    \label{fig:magnus-error-bound}
\end{figure}

\subsection{Magnus series convergence and truncation error}
Previously, we discussed the limitations of the Magnus expansion. Here, we
summarize its regime of validity, following \cite{Apel2025Magnus}.
Consider a general time-dependent Hamiltonian $H(t)$ defined on a finite
interval $t\in[0,t_f]$. We
define the maximal instantaneous energy scale as
\begin{equation}
h_{\mathrm{max}} \equiv \max_{t'\in[0,t_f]}\norm{ H(t')}\, ,
\end{equation}
where $\norm{\,\cdot\,}$ denotes the operator norm.
The convergence of the Magnus expansion is controlled by a universal
dimensionless constant $\delta_\xi$, defined as the inverse convergence radius,
$
\delta_\xi \equiv \xi^{-1}$, where $\xi$ is the universal convergence radius associated with the analytic structure of the Magnus series. The relevant
dimensionless expansion parameter is $\delta_\xi h_{\mathrm{max}} t$ and
sufficient condition for convergence is
\begin{equation}
\label{convergencebound}
\delta_\xi h_{\mathrm{max}} t < 1 \, .
\end{equation}
Within this regime, the Magnus series converges and provides a well-defined
logarithmic generator of the time-evolution operator.

For truncated dynamics, the operator-norm error between the exact Magnus
generator $\Omega(t)$ and its truncation $\Omega^{(N)}(t)$ is bounded as
\begin{equation}
\label{magnustruncationerror}
\norm{\Omega(t)-\Omega^{(N)}(t)}
\leq
\frac{4}{(N+1)^2}
\frac{\left(\delta_\xi h_{\mathrm{max}} t\right)^{N+1}}
{1-\delta_\xi h_{\mathrm{max}} t}\, .
\end{equation}
This bound shows that the truncation error is governed by the dimensionless
parameter $\delta_\xi h_{\mathrm{max}} t$, and decreases rapidly with increasing
truncation order $N$ as long as the evolution remains within the convergence
regime. As an illustration, we plot this error bound for the linearly polarized
Hamiltonian given in \eqref{eq:linpol-hamiltonian} in
Fig.~\ref{fig:magnus-error-bound}.

Because the global Magnus expansion has a finite convergence time, we
introduce a piecewise Magnus expansion. In this approach, the time evolution is divided into smaller subintervals such that the local convergence condition is satisfied on each step, allowing the effective Hamiltonian to be tracked over arbitrarily long times. Of course, one must account for the accumulation of local errors over many intervals; this will be addressed later on.

\section{From Global to Piecewise Magnus Expansion}

We now extend the discussion to systems that are not necessarily periodically driven. The goal is to develop a practical framework for extracting Krylov complexity from general time-dependent Hamiltonians. We focus on quadratic systems, for which the dynamics can be represented in terms of symplectic transformations. This setting is analytically tractable and also provides a useful testing ground for the piecewise Magnus method.

\subsection{Quadratic Oscillators, Symplectic Evolution, and Krylov Complexity}

Consider the harmonic oscillator with time-dependent frequency,
\begin{equation}
H_\mathrm{osc}(t)
=
\frac{p^2}{2m}
+
\frac{1}{2}m\omega^2(t)x^2 .
\end{equation}
Introducing the phase-space vector
\begin{equation}
\bm{R}
=
\begin{pmatrix}
x\\
p
\end{pmatrix},
\end{equation}
the Heisenberg equations can be written as
\begin{equation}
\dot{\bm{R}}(t)
=
M(t)\bm{R}(t),
\end{equation}
where
\begin{equation}
M(t)
=
\begin{pmatrix}
0 & 1/m\\
-m\omega^2(t) & 0
\end{pmatrix}.
\end{equation}
The formal solution is
\begin{equation}
\bm{R}(t)
=
S(t)\bm{R}(0),
\qquad
S(t)
=
\mathcal{T}
\exp\left[
\int_0^t M(s)\,ds
\right].
\end{equation}
Since \(M(t)\in\mathfrak{sp}(2,\mathbb{R})\), the propagator is symplectic,
\begin{equation}
S(t)\in Sp(2,\mathbb{R}).
\end{equation}
Writing the symplectic propagator as
\begin{equation}
S(t)
=
\begin{pmatrix}
A(t) & B(t)\\
C(t) & D(t)
\end{pmatrix},
\end{equation}
the symplectic condition implies
\begin{equation}
A(t)D(t)-B(t)C(t)=1.
\end{equation}

Now let us choose a reference oscillator frequency \(\omega_r>0\), and define
\begin{equation}
a
=
\sqrt{\frac{m\omega_r}{2}}\,x
+
\frac{i}{\sqrt{2m\omega_r}}\,p,
\qquad
a^\dagger
=
\sqrt{\frac{m\omega_r}{2}}\,x
-
\frac{i}{\sqrt{2m\omega_r}}\,p .
\end{equation}
Then the evolution induces the Bogoliubov transformation \cite{Weedbrook_2012}
\begin{equation}
U^\dagger(t,0)aU(t,0)
=
\mu(t)a+\nu(t)a^\dagger,
\end{equation}
where
\begin{equation}
\mu(t)
=
\frac{1}{2}
\left[
A(t)+D(t)
+
i\left(
\frac{C(t)}{m\omega_r}
-
m\omega_r B(t)
\right)
\right],
\end{equation}
and
\begin{equation}
\nu(t)
=
\frac{1}{2}
\left[
A(t)-D(t)
+
i\left(
\frac{C(t)}{m\omega_r}
+
m\omega_r B(t)
\right)
\right].
\end{equation}
These coefficients satisfy the canonical constraint
\begin{equation}
|\mu(t)|^2-|\nu(t)|^2=1.
\end{equation}
The squeezing parameter \(r(t)\) is defined by
\be
|\nu(t)|=\sinh r(t)\, .
\ee
If the initial state is the reference vacuum,
$
a\ket{0}=0
$, then quadratic evolution preserves Gaussianity, and the evolved state is a
squeezed vacuum. Its expansion in the Fock basis is\footnote{We neglect the squeezing phase, since it does not affect our results.}\cite{Weedbrook_2012},\cite{PhysRevA.13.2226}
\begin{equation}
\ket{\psi(t)}
=
\frac{1}{\sqrt{\cosh r(t)}}
\sum_{n=0}^{\infty}
\frac{\sqrt{(2n)!}}{2^n n!}
\left[\tanh r(t)\right]^n
\ket{2n}\, .
\end{equation}
Only even Fock states appear because
the squeezing operator creates and annihilates excitations in pairs.
Since the evolution from the reference vacuum remains in the even-parity sector,
a natural Krylov basis for this sector is
\begin{equation}
\ket{K_n}=\ket{2n},
\qquad
n=0,1,2,\dots .
\end{equation}
The probability distribution on the Krylov chain is
\begin{equation}
P_n(t)
=
\frac{(2n)!}{2^{2n}(n!)^2}
\frac{\tanh^{2n}r(t)}{\cosh r(t)}.
\end{equation}
And thus, the Krylov complexity reads
\begin{equation}
C_K(t)
=
\sum_{n=0}^{\infty} n P_n(t).
\end{equation}
Using either the generating function for \(P_n(t)\) or the relation to the mean
occupation number, one obtains
\begin{equation}
\label{OS}
C_K(t)
=
\frac{1}{2}\sinh^2 r(t)
=
\frac{1}{2}|\nu(t)|^2.
\end{equation}
The factor of \(1/2\) appears because the Krylov site \(n\) corresponds to the
physical Fock state \(\ket{2n}\). We have presented a detailed direct proof of this relation in Appendix \eqref{appC}.

\subsection{Global Magnus Benchmark: Sudden Quench}

Before introducing the fully piecewise construction, it is useful to examine a simple problem for which the usual global Magnus expansion can be applied directly. This occurs when the commutator structure is sufficiently simple; for example, when the relevant commutators vanish, or when the nested commutator algebra closes in a finite-dimensional Lie algebra. In such cases, the global Magnus expansion may terminate exactly or can be truncated in a controlled way over the full time interval.

This provides a useful analytical benchmark. After studying such a tractable case, we will be motivated to consider more complicated time-dependent systems for which a single global Magnus expansion is no longer convenient or accurate. For such problems, the piecewise Magnus expansion provides a more flexible approach: instead of approximating the entire time evolution by one global effective Hamiltonian, one divides the time interval into short segments and constructs a local effective Hamiltonian on each segment.

As a simple analytical benchmark, consider a sudden quench of the oscillator frequency,
\begin{equation}
\omega(t)
=
\begin{cases}
\omega_0, & t<0,\\
\omega_1, & t\geq 0,
\end{cases}
\end{equation}
with \(\omega_0,\omega_1>0\). For \(t>0\), the frequency is constant and thus in phase-space form,
\begin{equation}
\dot{\bm{R}}(t)=M_1\bm{R}(t),
\end{equation}
where
\begin{equation}
M_1
=
\begin{pmatrix}
0 & 1/m\\
-m\omega_1^2 & 0
\end{pmatrix}.
\end{equation}
Since \(M_1\) is time independent, all commutators appearing in the Magnus expansion vanish. Therefore, the global Magnus expansion terminates at first order
\begin{equation}
\Omega(t)
=
\Omega^{(1)}(t)
=
\int_0^t M_1\,ds
=
tM_1.
\end{equation}
Thus, the symplectic propagator is obtained exactly as
\begin{equation}
S(t)=e^{tM_1}.
\end{equation}
Using
\begin{equation}
M_1^2=-\omega_1^2 I_2,
\end{equation}
we find
\begin{equation}
S(t)
=
I_2\cos(\omega_1 t)
+
\frac{\sin(\omega_1 t)}{\omega_1}M_1.
\end{equation}
Therefore,
\begin{equation}
S(t)
=
\begin{pmatrix}
\cos(\omega_1 t)
&
\dfrac{\sin(\omega_1 t)}{m\omega_1}
\\[6pt]
-m\omega_1\sin(\omega_1 t)
&
\cos(\omega_1 t)
\end{pmatrix}.
\end{equation}
Taking the reference frequency to be \(\omega_r=\omega_0\), the Bogoliubov coefficient \(\nu(t)\) is
\begin{equation}
\nu(t)
=
\frac{i}{2}
\left(
\frac{\omega_0}{\omega_1}
-
\frac{\omega_1}{\omega_0}
\right)
\sin(\omega_1 t).
\end{equation}
Hence, for the even-Fock Krylov chain, \(\ket{K_n}=\ket{2n}\), the Krylov complexity can be obtained from Eq.~\eqref{OS}. The result is
\begin{equation}\label{sudden quench}
C_K(t)
=
\frac{1}{8}
\left(
\frac{\omega_0}{\omega_1}
-
\frac{\omega_1}{\omega_0}
\right)^2
\sin^2(\omega_1 t).
\end{equation}
This expression vanishes when \(\omega_1=\omega_0\), as expected in the absence of a quench.

This example illustrates a situation in which the global Magnus expansion is not merely approximate but exact, because the post-quench generator is time independent and the commutator hierarchy terminates immediately. However, for a genuinely time-dependent frequency \(\omega(t)\), the matrices \(M(t_1)\) and \(M(t_2)\) generally do not commute. In such cases, the global Magnus expansion over a long time interval may become complicated or inaccurate. This motivates the use of the piecewise Magnus expansion, in which the evolution is approximated locally on short time intervals.

\subsection{Piecewise Magnus Expansion for a General Time-Dependent Oscillator}

We now consider a more challenging analytical problem: the harmonic oscillator with a genuinely time-dependent frequency. In this case, the global Magnus expansion is generally difficult to evaluate in closed form, since the generators at different times do not, in general, commute. Consequently, a single effective generator over the full interval \([0,t]\) may either be analytically intractable or lie outside the convergence regime of the Magnus expansion. This motivates a local construction: we approximate the evolution over short time intervals, where the Magnus expansion is better controlled, and then compose the resulting local propagators.

To implement this idea, we partition the interval \([0,t]\) as
\begin{equation}
0=t_0<t_1<\cdots<t_N=t\, ,
\qquad
\Delta_j=t_{j+1}-t_j\, .
\end{equation}
The step sizes are chosen so that the Magnus expansion converges on each subinterval. From the convergence bound in \eqref{convergencebound}, this requires
$
\Delta_j < \frac{1}{\delta_\xi h_{\max}} 
$.
The advantage of the piecewise Magnus expansion is that the truncation error is controlled locally rather than over the full time interval. If the expansion is truncated at order \(m\), the global Magnus error scales as \(O(t^{m+1})\) within its convergence regime. By contrast, for a uniform partition with \(\Delta=t/N\), the local error on each subinterval scales as \(O(\Delta^{m+1})\). Assuming that these local errors accumulate at most linearly, up to stability constants, the total piecewise error is of order
\(O(N\Delta^{m+1})=O(t\Delta^m)\). Thus, by taking \(\Delta\) sufficiently small, the accumulated piecewise error can remain much smaller than the error of a single global Magnus expansion. Moreover, the piecewise method only requires the local convergence condition \(\delta_\xi h_{\max}\Delta<1\), so it can remain effective even when the global condition \(\delta_\xi h_{\max}t<1\) is violated.

In this method, the full symplectic propagator factorizes as
\begin{equation}
S(t)
=
S_{N-1}\cdots S_1S_0\, ,
\qquad
S_j\equiv S(t_{j+1},t_j)\, ,
\end{equation}
where, on each subinterval,
\begin{equation}
S_j=e^{\Omega_j},
\end{equation}
with
\begin{equation}
\Omega_j
=
\Omega_j^{(1)}
+
\Omega_j^{(2)}
+
\Omega_j^{(3)}
+\cdots .
\end{equation}
The first two terms are
\begin{equation}
\Omega_j^{(1)}
=
\int_{t_j}^{t_{j+1}}M(s)\,ds,
\end{equation}
and
\begin{equation}
\Omega_j^{(2)}
=
\frac{1}{2}
\int_{t_j}^{t_{j+1}}ds_1
\int_{t_j}^{s_1}ds_2\,
[M(s_1),M(s_2)]\, .
\end{equation}
Retaining terms up to second order gives
\begin{equation}
S_j
\approx
e^{\Delta_jM_{\mathrm{eff},j}},
\end{equation}
where
\begin{equation}
M_{\mathrm{eff},j}
=
\begin{pmatrix}
2\gamma_j & 1/m\\
-m\overline{\omega_j^2} & -2\gamma_j
\end{pmatrix}.
\end{equation}
Here
\begin{equation}
\overline{\omega_j^2}
=
\frac{1}{\Delta_j}
\int_{t_j}^{t_{j+1}}
\omega^2(s)\,ds
\end{equation}
is the interval-averaged squared frequency, and
\begin{equation}
\gamma_j
=
-\frac{1}{4\Delta_j}
\int_{t_j}^{t_{j+1}}ds_1
\int_{t_j}^{s_1}ds_2
\left[
\omega^2(s_2)-\omega^2(s_1)
\right] .
\end{equation}
The corresponding local effective Hamiltonian is
\begin{equation}
H_{\mathrm{eff},j}
=
\frac{p^2}{2m}
+
\frac{1}{2}m\overline{\omega_j^2}x^2
+
\gamma_j(xp+px).
\end{equation}
The total evolution is then approximated by multiplying the local symplectic steps in time order:
\begin{equation}
S(t)
\approx
e^{\Delta_{N-1}M_{\mathrm{eff},N-1}}
\cdots
e^{\Delta_1M_{\mathrm{eff},1}}
e^{\Delta_0M_{\mathrm{eff},0}} .
\end{equation}
This construction becomes increasingly accurate as the partition is refined. Once the approximate symplectic matrix \(S(t)\) has been obtained, the Bogoliubov coefficient \(\nu(t)\) follows from the formulas above, and the Krylov complexity is again read off from \eqref{OS}.

\subsection{Soft Quench}

To study a more realistic quench protocol and, at the same time, illustrate the
use of the piecewise Magnus expansion, let us consider a harmonic oscillator
with a smooth finite-time frequency quench. In this protocol, the oscillator
frequency changes continuously from its initial value \(\omega_0\) to its final
value \(\omega_1\) over a finite time scale \(\tau\). We choose
\begin{equation}
\omega^2(t)
=
\begin{cases}
\omega_0^2,
&
t<0,
\\[4pt]
\omega_0^2
+
\left(\omega_1^2-\omega_0^2\right)
\sin^2\left(\dfrac{\pi t}{2\tau}\right),
&
0\leq t\leq \tau,
\\[8pt]
\omega_1^2,
&
t>\tau .
\end{cases}
\end{equation}
Here, \(\tau\) controls the duration of the quench. In the limit
\(\tau\to 0\), this protocol approaches the sudden quench, whereas for large
\(\tau\) the evolution becomes increasingly close to an adiabatic change of the
oscillator frequency~\cite{RevModPhys.83.863, Zurek2005}. This profile
is continuous and has a vanishing first derivative at the beginning and end of
the ramp. It therefore avoids the discontinuity of the sudden quench and provides
a natural setting in which to test the piecewise Magnus expansion.

\begin{figure}[h]
    \centering
    \includegraphics[width=0.7\textwidth]{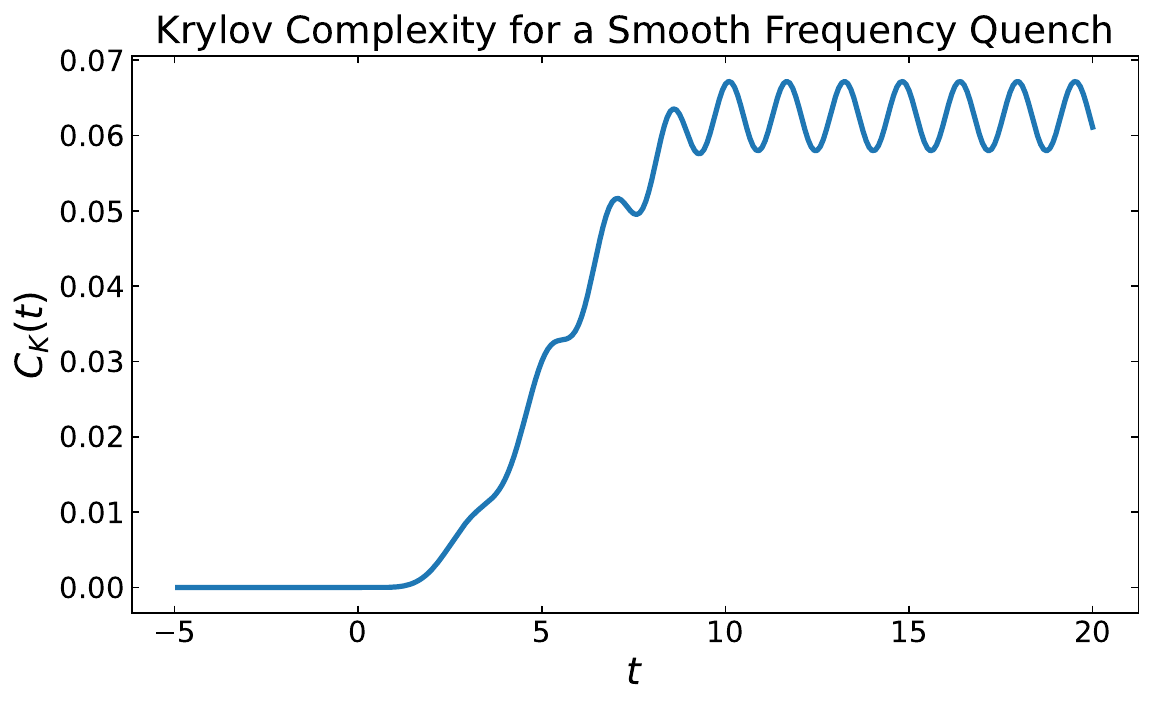}
    \caption{Krylov complexity for a soft quench with parameters
    \(\omega_0=1\), \(\omega_1=2\), and \(\tau=10\).}
    \label{fig1}
\end{figure}

Applying the piecewise Magnus algorithm, we compute the symplectic propagator
numerically, extract the corresponding Bogoliubov coefficient \(\nu(t)\), and
then obtain the Krylov complexity from \eqref{OS}. The resulting behavior
depends strongly on the quench duration \(\tau\). For a moderately fast smooth
quench, the complexity grows during the ramp and then exhibits bounded
oscillations after the quench is completed, as shown in Fig.~\ref{fig1}. This
behavior is expected: during the time-dependent ramp, the state is driven away
from the initial vacuum, while after the ramp the system evolves with the final
constant frequency \(\omega_1\), producing oscillations associated with the
resulting squeezed state.

\begin{figure}[h]
    \centering
    \includegraphics[width=0.7\textwidth]{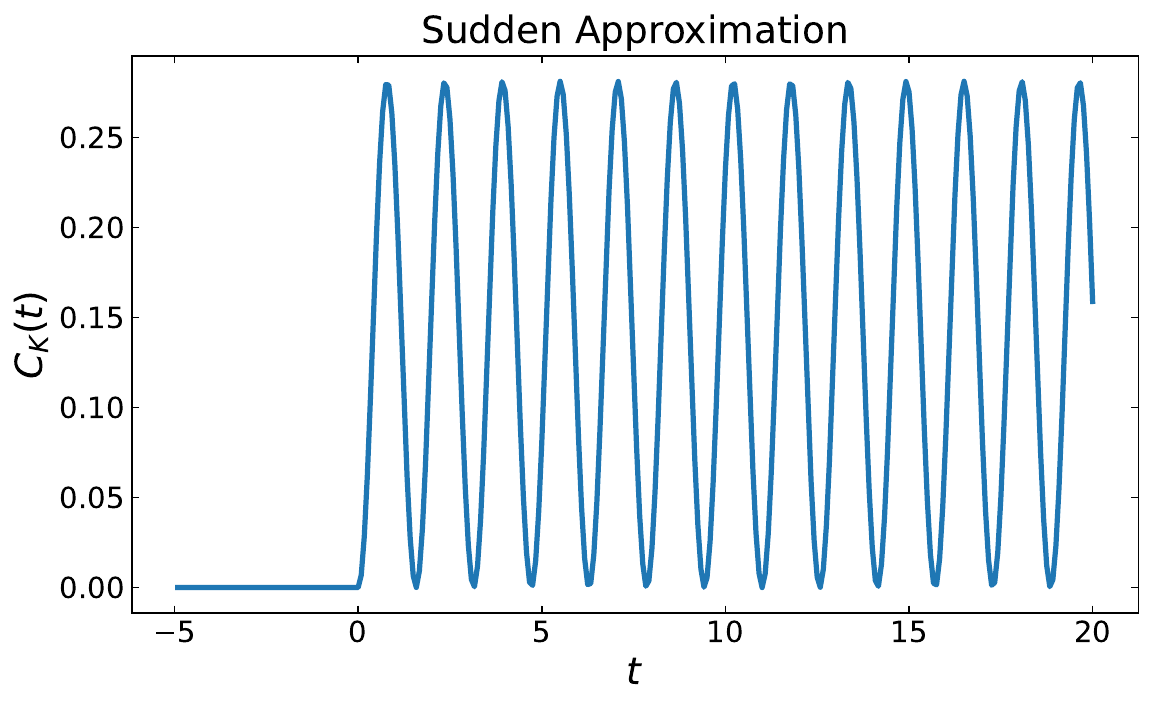}
    \caption{Krylov complexity for an almost sudden quench with parameters
    \(\omega_0=1\), \(\omega_1=2\), and \(\tau=0.1\).}
    \label{fig2}
\end{figure}

For a very rapid quench, the result approaches the sudden-quench limit. In this
case, the numerical curve obtained from the piecewise Magnus method agrees well
with the exact analytical expression derived earlier for the sudden quench; see \eqref{sudden quench}. This is illustrated in Fig.~\ref{fig2}.

\begin{figure}[h]
    \centering
    \includegraphics[width=0.7\textwidth]{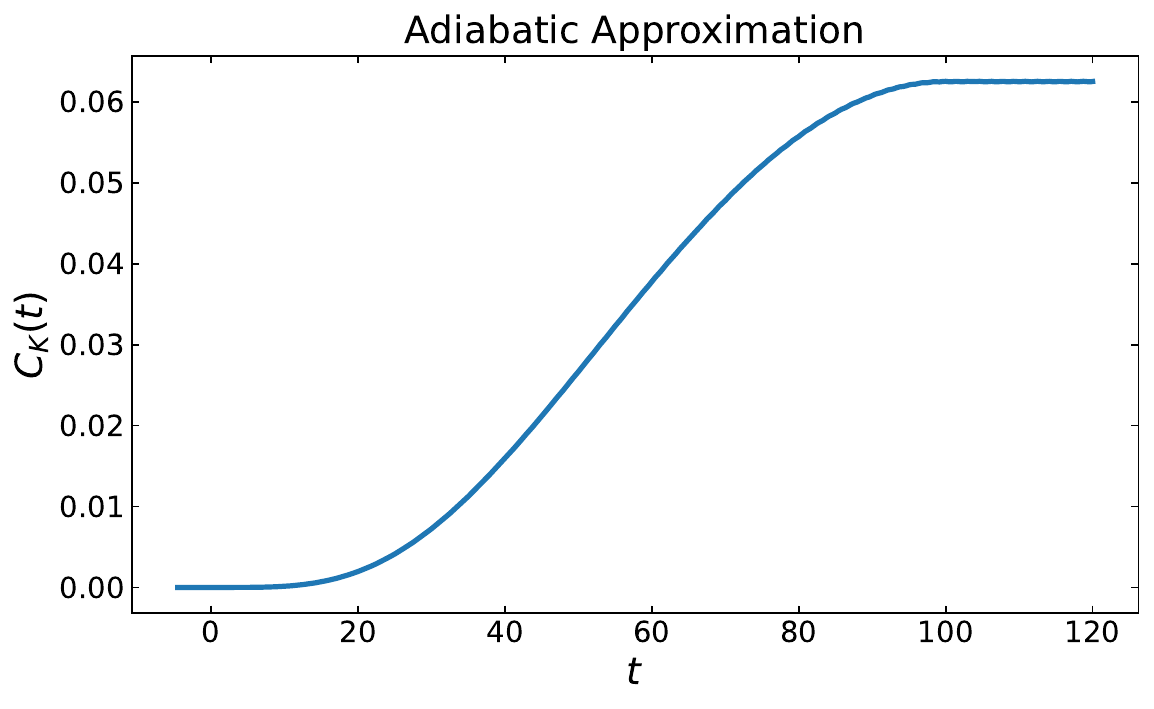}
    \caption{Krylov complexity for a very slow soft quench with parameters
    \(\omega_0=1\), \(\omega_1=2\), and \(\tau=100\).}
    \label{fig3}
\end{figure}

On the other hand, for a very slow quench, the evolution becomes approximately
adiabatic. In this regime, the system closely follows the instantaneous ground
state of the Hamiltonian. The corresponding complexity increases smoothly and
approaches a nearly constant final value, as shown in Fig.~\ref{fig3}. The
absence of sizable post-ramp oscillations indicates that only a small amount of
nonadiabatic excitation is produced.

These behaviors are consistent with physical expectations and agree with the
known analytical results in the appropriate limits. They therefore support the
reliability of the piecewise Magnus expansion in the present setting. This method
can thus serve as a useful tool for studying more general time-dependent
quadratic systems, and it would be interesting to test it for a broader class of
time-dependent Hamiltonians.
\section{Conclusion}

In this work, we investigated the Krylov complexity of states evolving under time-dependent Hamiltonians. Our analysis was carried out analytically whenever possible. We first considered periodically driven systems, for which Floquet theory provides a natural framework. In this setting, we used Floquet's theorem to obtain the stroboscopic complexity at integer multiples of the driving period.

For systems in which the exact Floquet Hamiltonian cannot be obtained
analytically in closed form, we employed the Magnus expansion as a systematic approximation scheme. In particular, for the linearly polarized periodically driven two-level system, we showed that the resulting high-frequency expansion can be expressed in terms of special functions, namely Bessel and Struve functions.

We then turned to more general, non-periodic time-dependent potentials, for
which Floquet theory is not applicable. In this broader setting, we argued that the Magnus expansion remains a useful methodology. As a concrete example, we studied the harmonic oscillator with a time-dependent frequency. For cases in which the commutator structure is sufficiently simple, such as the sudden quench, the global Magnus expansion can be evaluated exactly. For a generic time-dependent frequency profile, however, a single global Magnus expansion is generally no longer convenient or sufficiently accurate over long time intervals.

This observation motivated the use of the piecewise Magnus expansion. We applied this method to a smooth finite-time quench protocol and computed the corresponding Krylov complexity using the proposed algorithm. The resulting behavior agreed with physical expectations, and in the appropriate limiting cases the method reproduced known analytical results. These findings support the use of the piecewise Magnus expansion as a practical and reliable tool for studying Krylov complexity in a broader class of time-dependent systems.

It is useful to place our approach in the context of earlier attempts to extend Krylov methods to time-dependent problems. An important development, which also addresses Floquet evolution in the context of operator Krylov complexity, is the
work of \cite{Yeh:2023fek}. There, it was shown that the stroboscopic
evolution of any Hermitian operator under a Floquet unitary can be mapped to an effective one-dimensional non-interacting problem in Krylov space, namely a Floquet transverse-field Ising model with inhomogeneous couplings. In this description, the full Floquet Krylov structure is encoded in a set of Krylov angles, which determine the effective transverse fields and Ising couplings of the mapped model. This mapping is particularly significant because it demonstrates that Floquet Krylov dynamics admits a universal description, independent of the spatial dimension or the local Hilbert-space structure of the original system. It further suggests that autocorrelation functions, and more generally operator dynamics in Floquet systems, can be understood in terms of the topological and dynamical properties of this effective non-interacting model.

Within Floquet theory, related approaches were developed in \cite{Yates2021StrongModesKrylov,Nizami2023KrylovFloquet,Nizami2024SpreadComplexity},
where the Krylov basis is generated using the Arnoldi iteration. In that
framework, the dynamics are mapped onto a one-dimensional chain with long-range hopping, in contrast to the conventional Lanczos-based Krylov construction, which yields only nearest-neighbor hopping along the Krylov chain.

An alternative method for treating time-dependent Hamiltonians is the extended
Hilbert-space formalism developed in \cite{Takahashi:2024hex}. In this approach, one introduces an auxiliary
evolution parameter \(s\) and treats the physical time \(t\) as an additional
coordinate. The evolution is defined by
\begin{equation}
\ket{\psi(s,t)}
=
e^{-is\left(H(t)-i\partial_t\right)}
\ket{\phi(t)},
\end{equation}
where \(\ket{\phi(t)}\) is a reference state satisfying
\begin{equation}
\ket{\phi(0)}=\ket{\psi(0)}.
\end{equation}
The physical time-evolved state is then recovered by setting
\begin{equation}
\ket{\psi(t)}=\ket{\psi(s,t)}\big|_{s=t}.
\end{equation}
In this formalism, the operator \(H(t)-i\partial_t\) acts as an effective
time-independent Hamiltonian in an enlarged Hilbert space. This construction
makes it possible to apply the standard Krylov method to time-dependent
generators while preserving the one-dimensional lattice representation of the
dynamics. In particular, one may construct the Krylov basis from the set of
vectors
\begin{equation}
\left\{
\left(H(t)-i\partial_t\right)^n\ket{\phi(t)}
\right\}_{n=0}^{\infty},
\end{equation}
thereby reformulating the original time-dependent evolution in terms of a
static generator in the extended space.

A related and more recent development is the use of the diabatic Magnus
expansion to study Krylov complexity in explicitly time-dependent settings,
especially across quantum phase transitions~\cite{Grabarits2025}. In this
approach, one factors out the adiabatic evolution and defines a diabatic
time-evolution operator that captures only the nonadiabatic part of the
dynamics. This framework is particularly useful when one is interested in the
complexity generated purely by departures from adiabaticity. Using the
transverse-field Ising model as a testbed, it was shown that the growth of
Krylov complexity across a quantum critical point exhibits universal
Kibble--Zurek scaling \cite{Zurek2005,delCampo2014}. In particular, the occupation probabilities in Krylov
space take a Poissonian form in the scaling regime, and the cumulants of the
complexity display universal power-law behavior. These results suggest that
Krylov complexity can serve as a useful probe of nonequilibrium universality in
driven many-body systems. For other investigations of Krylov complexity in
time-dependent problems, see \cite{Adolfo,Alishahiha}.

We believe that our work complements these efforts by providing a concrete
Magnus-based framework for studying how complexity spreads in systems governed
by dynamical Hamiltonians. As future directions, it would be natural to explore
additional explicit examples and to extend our strategy to the study of
operator complexity. Another interesting direction is to consider generalized
Krylov complexity~\cite{FarajiAstaneh:2025thi} and investigate systems whose
Hamiltonians depend explicitly on transformation parameters rather than on time
alone. We leave these questions for future work.

\section*{Acknowledgments}

We thank Mohsen Alishahiha and Adolfo del Campo for reading our draft and for
their useful comments.

\newpage
\appendix
\counterwithin{equation}{section}
\section{A Review on Krylov Complexity for Time-Independent Hamiltonians}\label{appA}
Before considering explicitly time-dependent systems, we briefly review the state Krylov construction for a time-independent Hamiltonian \(H\), which could also be an effective Hamiltonian \(H_{\mathrm{eff}}\) derived from a time-dependent Hamiltonian \(H(t)\) \cite{Parker:2018yvk,Balasubramanian:2022hcx}. Given an initial state \(\ket{\psi_0}\), the first Krylov vector is defined as
\begin{equation}
\ket{K_0}=\ket{\psi_0}.
\end{equation}
The Krylov basis is generated by the Lanczos recursion. Starting from \(\ket{K_0}\), define
\begin{equation}
a_n=\bra{K_n}H\ket{K_n}.
\end{equation}
The residual vector is
\begin{equation}
\ket{w_n}
=
H\ket{K_n}
-
a_n\ket{K_n}
-
b_n\ket{K_{n-1}},
\end{equation}
with the initial convention
\begin{equation}
b_0=0,
\qquad
\ket{K_{-1}}=0.
\end{equation}
The next Lanczos coefficient and Krylov vector are then
\begin{equation}
b_{n+1}=\norm{\ket{w_n}},
\end{equation}
and
\begin{equation}
\ket{K_{n+1}}
=
\frac{1}{b_{n+1}}\ket{w_n},
\end{equation}
provided \(b_{n+1}\neq 0\). This construction gives an orthonormal Krylov basis
\begin{equation}
\{\ket{K_0},\ket{K_1},\ket{K_2},\dots\}.
\end{equation}
Expanding the time-evolved state as
\begin{equation}
\ket{\psi(t)}
=
\sum_{n=0}^{N_K-1}\phi_n(t)\ket{K_n},
\end{equation}
the state Krylov complexity, or spread complexity, is then defined by
\begin{equation}
C_K(t)
=
\sum_{n=0}^{N_K-1}
n|\phi_n(t)|^2.
\end{equation}
This construction will be applied whenever the time-dependent problem can be mapped to an effective time-independent Hamiltonian, either exactly or approximately.

\section{Derivation of the Second Dressed-Frame Magnus Term}
\label{appB}

In this appendix we derive the second Magnus correction used in
Sec.~\eqref{section4.2.2} for the linearly polarized two-level system in
the dressed frame. The dressed-frame Hamiltonian is
\begin{equation}
\widetilde{H}_{\mathrm{lin}}(t)
=
a\left[
\cos\theta(t)\,\sigma_z+\sin\theta(t)\,\sigma_y
\right]\, ,
\qquad
a\equiv\frac{\omega_0}{2}\, ,
\qquad
\theta(t)\equiv b\sin(\omega t)\, ,
\qquad
b\equiv\frac{2\Omega_0}{\omega}\, .
\end{equation}
The second Magnus contribution to the effective Floquet Hamiltonian is
\begin{equation}
H_{\mathrm{eff}}^{(2)}
=
-\frac{i}{2T}
\int_0^T dt_1
\int_0^{t_1}dt_2\,
[
\widetilde{H}_{\mathrm{lin}}(t_1),
\widetilde{H}_{\mathrm{lin}}(t_2)
],
\qquad
T=\frac{2\pi}{\omega}.
\end{equation}
Using the Pauli algebra, one finds
\begin{align}
[
\widetilde{H}_{\mathrm{lin}}(t_1),
\widetilde{H}_{\mathrm{lin}}(t_2)
]
&=
2ia^2
\sin\!\left[
\theta(t_1)-\theta(t_2)
\right]\sigma_x .
\end{align}
Therefore,
\begin{equation}
H_{\mathrm{eff}}^{(2)}
=
\frac{a^2}{T}
\int_0^T dt_1
\int_0^{t_1}dt_2\,
\sin\!\left[
\theta(t_1)-\theta(t_2)
\right]\sigma_x .
\end{equation}
Introducing dimensionless phase variables
\be
\varphi_{1,2}=\omega\, t_{1,2}\, ,
\ee
we obtain
\begin{equation}
H_{\mathrm{eff}}^{(2)}
=
\frac{a^2}{T\omega^2}
\mathcal I(b)\,\sigma_x,
\end{equation}
where
\begin{equation}
\mathcal I(b)
=
\int_0^{2\pi} d\varphi_1
\int_0^{\varphi_1} d\varphi_2\,
\sin\!\left[
b\sin \varphi_1-b\sin \varphi_2
\right].
\end{equation}
We write
\begin{equation}
\mathcal I(b)
=
\mathcal I_1(b)-\mathcal I_2(b),
\end{equation}
with
\begin{align}
\mathcal I_1(b)
&=
\int_0^{2\pi} d\varphi_1\,
\sin(b\sin \varphi_1)
\int_0^{\varphi_1} d\varphi_2\,
\cos(b\sin \varphi_2),
\\
\mathcal I_2(b)
&=
\int_0^{2\pi} d\varphi_1\,
\cos(b\sin \varphi_1)
\int_0^{\varphi_1} d\varphi_2\,
\sin(b\sin \varphi_2).
\end{align}
We now use the standard Bessel-function expansions
\begin{align}
\cos(b\sin \varphi)
&=
J_0(b)
+
2\sum_{r=1}^{\infty}
J_{2r}(b)\cos(2r\varphi),
\\
\sin(b\sin \varphi)
&=
2\sum_{\ell=0}^{\infty}
J_{2\ell+1}(b)\sin[(2\ell+1)\varphi].
\end{align}
First consider \(\mathcal I_1\). Since
\begin{equation}
\int_0^{\varphi_1} d\varphi_2\,
\cos(b\sin \varphi_2)
=
J_0(b)\varphi_1
+
2\sum_{r=1}^{\infty}
\frac{J_{2r}(b)}{2r}\sin(2r\varphi_1),
\end{equation}
and since the integral over a product of an odd sine and an even sine vanishes
over \([0,2\pi]\), only the term proportional to
\(J_0(b)\varphi_1\) contributes. Thus
\begin{align}
\mathcal I_1(b)
&=
J_0(b)
\int_0^{2\pi} d\varphi_1\,
\varphi_1\,\sin(b\sin \varphi_1)
\nonumber\\
&=
2J_0(b)
\sum_{\ell=0}^{\infty}
J_{2\ell+1}(b)
\int_0^{2\pi} d\varphi_1\,
\varphi_1\sin[(2\ell+1)\varphi_1].
\end{align}
Using
\begin{equation}
\int_0^{2\pi} d\varphi\,\varphi\sin(n\varphi)
=
-\frac{2\pi}{n},
\qquad
n\in\mathbb{Z}_+,
\end{equation}
we obtain
\begin{equation}
\mathcal I_1(b)
=
-4\pi J_0(b)
\sum_{\ell=0}^{\infty}
\frac{J_{2\ell+1}(b)}{2\ell+1}.
\end{equation}
The series can be expressed in terms of the Struve function
\(\mathbf H_0\)\cite{NIST:DLMF}
\begin{equation}
\sum_{\ell=0}^{\infty}
\frac{J_{2\ell+1}(b)}{2\ell+1}
=
\frac{\pi}{4}\mathbf H_0(b).
\end{equation}
Therefore
\be
\mathcal I_1(b)=-\pi^2J_0(b)\mathbf H_0(b)\, .
\ee
Similarly, for \(\mathcal I_2\), we get exactly the same result with a negative sign.
Thus,
\begin{equation}
H_{\mathrm{eff}}^{(2)}
=
-\frac{\pi a^2}{\omega}
J_0(b)\mathbf H_0(b)\sigma_x.
\end{equation}
Restoring the original parameters, we finally obtain
\begin{equation}
H_{\mathrm{eff}}^{(2)}
=
-\frac{\pi\omega_0^2}{4\omega}
J_0\!\left(\frac{2\Omega_0}{\omega}\right)
\mathbf H_0\!\left(\frac{2\Omega_0}{\omega}\right)
\sigma_x.
\end{equation}

\section{Direct Derivation of the Squeezed-Vacuum Krylov Complexity}\label{appC}

For completeness, we provide a direct derivation of the Krylov complexity of the squeezed vacuum from its probability distribution. The squeezed-vacuum state has the expansion
\begin{equation}
\ket{\psi(t)}
=
\frac{1}{\sqrt{\cosh r(t)}}
\sum_{n=0}^{\infty}
\frac{\sqrt{(2n)!}}{2^n n!}
\left[
\tanh r(t)
\right]^n
\ket{2n}.
\end{equation}
Choosing the even-Fock Krylov chain
\begin{equation}
\ket{K_n}=\ket{2n},
\qquad
n=0,1,2,\dots,
\end{equation}
the probability of occupying the \(n\)-th Krylov site is
\begin{equation}
P_n(t)
=
\frac{(2n)!}{2^{2n}(n!)^2}
\frac{\tanh^{2n}r(t)}{\cosh r(t)}.
\end{equation}
The Krylov complexity is
\begin{equation}
C_K(t)
=
\sum_{n=0}^{\infty}nP_n(t).
\end{equation}
Let
\begin{equation}
x=\tanh^2 r(t)\, ,
\end{equation}
then
\begin{equation}
C_K(t)
=
(1-x)^{1/2}
\sum_{n=0}^{\infty}
n
\frac{(2n)!}{2^{2n}(n!)^2}
x^n.
\end{equation}
Using the generating function
\begin{equation}
\sum_{n=0}^{\infty}
\frac{(2n)!}{2^{2n}(n!)^2}
x^n
=
\frac{1}{\sqrt{1-x}},
\end{equation}
we obtain
\be
\sum_{n=0}^{\infty}
n
\frac{(2n)!}{2^{2n}(n!)^2}
x^n
=
x\frac{d}{dx}
\left[
\frac{1}{\sqrt{1-x}}
\right]
\nonumber
=
\frac{x}{2(1-x)^{3/2}}.
\ee
Therefore,
\begin{equation}
C_K(t)
=\frac{x}{2(1-x)}.
\end{equation}
Substituting for $x$ we find
\begin{equation}
C_K(t)
=
\frac{1}{2}\sinh^2 r(t).
\end{equation}
Using
\begin{equation}
|\nu(t)|=\sinh r(t),
\end{equation}
this becomes
\begin{equation}
C_K(t)
=
\frac{1}{2}|\nu(t)|^2.
\end{equation}
\bibliographystyle{unsrt}
\bibliography{ref}
\end{document}